\documentclass[sigconf]{acmart}
\usepackage{braket}

\usepackage{multirow}
\AtBeginDocument{%
  \providecommand\BibTeX{{%
    \normalfont B\kern-0.5em{\scshape i\kern-0.25em b}\kern-0.8em\TeX}}}

\copyrightyear{2021}
\acmYear{2021}
\setcopyright{acmcopyright}
\acmConference[KDD '21] {Proceedings of the 27th ACM SIGKDD Conference on Knowledge Discovery and Data Mining}{August 14--18, 2021}{Virtual Event, Singapore.}
\acmBooktitle{Proceedings of the 27th ACM SIGKDD Conference on Knowledge Discovery and Data Mining (KDD '21), August 14--18, 2021, Virtual Event, Singapore}
\acmPrice{15.00}
\acmISBN{978-1-4503-8332-5/21/08}
\acmDOI{10.1145/3447548.3467151}

\newcounter{tightlistcnt}
\newenvironment{tightitemize}{%
    \begin{list}{\textbullet}{\usecounter{tightlistcnt}%
    \topsep 0in
    \partopsep 0in
    \itemsep .2em
    \parsep 0in
    \leftmargin 0em
    \rightmargin 0in
    \listparindent 0em
    \itemindent .75em
    \labelsep .75em
    \labelwidth 0in
    }%
}{%
    \end{list}%
}

\newenvironment{tightenumerate}{%
    \begin{list}{\arabic{tightlistcnt}.}{\usecounter{tightlistcnt}
    \topsep 0in
    \partopsep 0in
    \itemsep .2em
    \parsep 0in
    \leftmargin 0.0em
    \rightmargin 0in
    \listparindent 0em
    \itemindent .75em
    \labelsep .75em
    \labelwidth 0in
    }%
}{%
    \end{list}%
}

\begin{document}

\fancyhead{}
\title{Leveraging Tripartite Interaction Information from Live Stream E-Commerce for Improving Product Recommendation}





\author{Sanshi Yu$^1$\footnotemark[1], Zhuoxuan Jiang$^{2}$\footnotemark[1], Dong-Dong Chen$^3$, Shanshan Feng$^4$, \\Dongsheng Li$^5$, Qi Liu$^1$, Jinfeng Yi$^3$}

\affiliation{
 \institution{$^1$University of Science and Technology of China, Hefei, China}
 \country{}
}
\affiliation{
 \institution{$^2$Tencent, Shanghai, China}
 \country{}
}
\affiliation{
 \institution{$^3$JD AI Research, Beijing, China}
 \country{}
}
\affiliation{
 \institution{$^4$Harbin Institute of Technology, Shenzhen, China}
 \country{}
}
\affiliation{
 \institution{$^5$Microsoft Research Asia, Shanghai, China}
 \country{}
}

\email{jzhx@pku.edu.cn, meet.leiyu@gmail.com, {chendongdong1, yijinfeng}@jd.com, }
\email{victor_fengss@foxmail.com, dongshengli@fudan.edu.cn, qiliuql@ustc.edu.cn}

\begin{abstract}
Recently, a new form of online shopping becomes more and more popular, which combines live streaming with E-Commerce activity. The streamers introduce products and interact with their audiences, and hence greatly improve the performance of selling products. Despite of the successful applications in industries, the live stream E-commerce has not been well studied in the data science community.
To fill this gap, we investigate this brand-new scenario and collect a real-world Live Stream E-Commerce (LSEC) dataset. Different from conventional E-commerce activities, the streamers play a pivotal role in the LSEC events. Hence, the key is to make full use of rich interaction information among streamers, users, and products. 
We first conduct data analysis on the tripartite interaction data and quantify the \textbf{\textit{streamer's influence}} on users' purchase behavior. Based on the analysis results, we model the tripartite information as a heterogeneous graph, which can be decomposed to multiple bipartite graphs in order to better capture the influence. We propose a novel Live Stream E-Commerce Graph Neural Network framework (LSEC-GNN) to learn the node representations of each bipartite graph, and further design a multi-task learning approach to improve product recommendation. Extensive experiments on two real-world datasets with different scales show that our method can significantly outperform various baseline approaches.

\end{abstract}



\maketitle

\renewcommand{\thefootnote}{\fnsymbol{footnote}} 
\footnotetext[1]{The authors contributed equally to this work, which was done when they were in JD AI Research and Sanshi Yu was an intern. Zhuoxuan Jiang and Shanshan Feng are the corresponding authors.} 


\section{Introduction}
\begin{figure}
  \centering
  \includegraphics[width=0.8\linewidth]{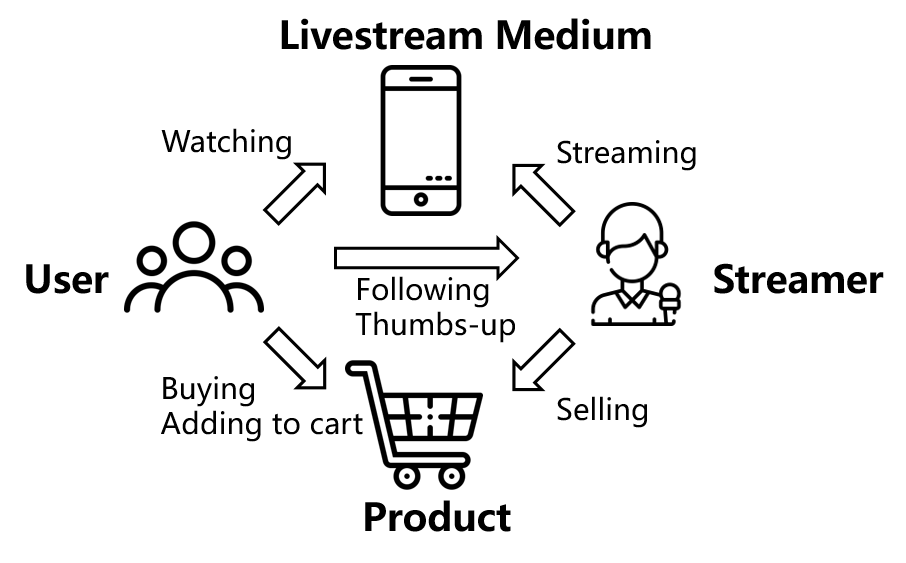}
  \vspace{-2ex}
  \caption{Example of the live stream E-Commerce scenario with the interactions among users, streamers and products.}
  \label{fig:stream}
  \vspace{-4ex}
\end{figure}

Recent years witness the prosperity of online live streaming. With the development of mobile phones, cameras, and high-speed internet, more and more users are able to broadcast their experiences in live streams on various social platforms, such as Facebook Live and YouTube Live.
There are a variety of live streaming applications~\cite{CHI2018}, including knowledge share, video-gaming, and outdoor traveling. One of the most important scenarios is live streaming commerce, where streamers broadcast themselves online and promote products to their audiences, as shown in Figure~\ref{fig:stream}.

This new form of sales can greatly shorten the decision-making time of consumers and provoke the sales volume~\cite{Hawaii2019,hu2020enhancing}. For example, according to a report from Alibaba’s Taobao Live platform, there are several numbers worthy of attention in last year’s 11.11 Global Shopping Festival: (1) the total GMV driven by live streaming achieved \$6 Billion USD, (2) approximately 300 million Taobao users watched live streams, and (3) 33 live streaming channels achieved over \$15 million USD in sales and nearly 500 live streaming channels reached \$1.5 million USD~\cite{report}. Some quantitative research results show that adopting live streaming in sales can achieve a 21.8\% increase in online sales volume~\cite{Hawaii2019}.


This new kind of shopping experience is significantly different from conventional online shopping, where only static information, e.g., images and texts, are available for customers. The expert streamers introduce and promote the products in a live streaming manner, which makes the shopping process more interesting and convincing.  In addition, another feature of live streaming is the rich and real-time interactions between streamers and their 
audiences, which makes live streaming a new medium and a powerful marketing tool for E-Commerce. Compared with the traditional TV live shopping, viewers not only can watch the showing for product's looks and functions, but also can ask the streamers to show different or individual perspectives of the products in real-time.

Fortunately, all the interaction information is recorded, such as watching behavior, thumbs-up, sending a message, and placing an order. Some studies by data analysis have found that some viewers are directly influenced by the streamers and may trust the streamers' recommendations.
The viewers could also indirectly influence other viewers due to their similar purchase preference~\cite{CHI2018}.

Despite the above finding that the streamers have the influence to stimulate consumers' purchase behavior, 
some more detailed and quantitative questions are still less-studied. 
For example, a pivotal question is that how and to what degree do the streamers influence the users' purchase behavior? In this paper, we investigate the question through data analysis, and find three patterns that the streamers can make an impact on users. We name it as \textbf{\textit{streamer's influence}} and summarize the three patterns in the following: 

\begin{enumerate}
    \item \textbf{The streamer can connect users with items}:  We model the probability of purchase for the users in two experiment settings and find that, the users are 4.9 times more probable to buy the products that are promoted by their followed streamers than those who buy the products that are not promoted by their followed streamers.
    \item \textbf{The streamer can connect similar users}: We introduce a similarity metric to calculate the similarity of purchased products among users in two experiment settings and find that, the average similarity of purchased products for two users who follow the same streamer is 4.6 times larger than that for two users who do not follow the same streamer.
    \item \textbf{The streamer can connect similar products}: Similar to the previous experiment, we introduce a similarity metric to calculate the similarity of user groups on products in two experiment settings and find that, on average, the user groups who purchase products recommended by the same streamer is 4.5 times more similar than user groups who purchase products which are not recommended by the same streamer.
\end{enumerate}

The analysis results suggest that the \textbf{\textit{streamer's influence}} can be captured through the interaction information to model the users' purchase behavior. More detailed data analysis can be referred to Section 4.



Based on our above findings, we propose a novel Live Stream E-Commerce Graph Neural Network (LSEC-GNN) framework to improve the classical product recommendation task. The tripartite interaction information is modeled as a heterogeneous graph, which is composed of multiple bipartite graphs in order to better capture the mutual influence. Then the LSEC-GNN learns and aggregates the node representations of each bipartite graph to predict product purchase behaviors by multi-task learning. Note that there may be diverse kinds of interactions between two nodes and our framework is flexible enough. To the best of our knowledge, there are no public live stream datasets. We collect two novel datasets from 
an E-Commerce website with different scales. The experimental results show that our LSEC-GNN is effective to improve the product recommendation task by thorough ablation and visualization studies.


To summarize, we make the following contributions:

\begin{tightitemize}
   
    \item We conduct thorough data analysis for quantifying how and to what degree the streamers make an impact on users' purchase behavior.
    \item We identify the problem of product recommendation under the new scenario of live stream E-Commerce by leveraging tripartite interaction information. And the \textbf{\textit{streamers' influence}} is first proposed. To the best of our knowledge, we are the first to study the recommendation for live streaming shopping. 
    \item We propose a Live Stream E-Commerce Graph Neural Network (LSEC-GNN) framework for modeling the tripartite interaction information as a heterogeneous graph and improving product recommendation by multi-task learning.
    \item We collect two new real-world datasets of tripartite interactions from an E-Commerce website. The experiment results can support our findings and our method can significantly outperform various baseline approaches. 
    \item We publish the real-world live stream datasets and code of our models to facilitate the community for future research works in this emerging area (\url{https://github.com/yusanshi/LSEC-GNN}).
\end{tightitemize}

\section{Related Work}


Online live streaming has  drawn increasing attention recently in social science~\cite{CHI2018,HILVERTBRUCE201858,Scheibe2016}, marketing science~\cite{SUN2019100886,hu2020enhancing}, and computer science~\cite{shen2018organizing}, etc. User activities in online live streaming platforms have been extensively studied including motivation~\cite{CHI2018}, social and community interaction~\cite{HILVERTBRUCE201858}, information production behavior~\cite{Scheibe2016}, channel selection~\cite{shen2018organizing}, etc. Besides, studies have shown that user online shopping behaviors may be different in online live streaming platforms. Chen et al.~\cite{Hawaii2019} reported a 21.8\% increase in online sales volume after adopting the live streaming strategy. Sun et al.~\cite{SUN2019100886} found that visibility affordance, metavoicing affordance, and guidance shopping affordance are the most influential factors on customer purchase intentions. Hu et al.~\cite{hu2020enhancing} found that social and structural bonds positively affect consumer engagement, while financial bonds have only an indirect effect. However, to the best of our knowledge, there is no existing work to study the recommendation problem in live stream E-Commerce. 

Our work is mainly related to product recommendation with user-item interaction information. Traditionally, matrix factorization (MF) methods achieved huge success in both rating prediction tasks~\cite{koren2009matrix,BPMF,lee2013local,li2017mrma} and top-N recommendation tasks~\cite{Hu08,Rendle09}. Recently, neural network based methods~\cite{Sedhain15,NCF,HidasiICLR16,rrn2017,liang2018variational} further improved the performance of product recommendation, especially in the top-N recommendation setting. For instance, the NCF method~\cite{NCF} improved the traditional MF methods by introducing MLPs to model the non-linear interactions between users and items. Similarly, many advanced neural network methods were applied in recommendation tasks to further enhance the state-of-the-art performance, e.g., autoencoders~\cite{Sedhain15,liang2018variational}, recurrent neural networks~\cite{HidasiICLR16,rrn2017}, etc. More recently, graph neural networks were introduced to the many recommendation tasks, which can improve the performance by combining both user rating information and the user-item interaction graph information~\cite{NGCF,LightGCN,GC-MC,HashGNN,MCCF,LR-GCCF,IG-MC,Bi-HGNN,STAR-GCN}. However, these GNN based methods were focused on bipartite graphs, which may not be optimal for tripartite graphs in product recommendation problem of live stream E-Commerce.

Similar to this work, many existing works incorporated additional graphs beside the user-item interaction graphs in recommendation tasks, e.g., knowledge graphs~\cite{KGNN-LS,KGCN,KGAT,AKGE} and social graphs~\cite{DANSER,GraphRec,DiffNet,Yu2020,song2019session}. For instance, Wang et al.~\cite{KGNN-LS} proposed the knowledge-aware graph neural network with label smoothness regularization, which first applies a trainable
function to identify important KG relationships and then applies a GNN to compute personalized item embeddings. Wang et al.~\cite{KGAT} proposed the KGAT method, which can capture the high-order relations from the hybrid structure of KG and user-item graph via attention mechanism. Wu et al.~\cite{DANSER} proposed a dual graph attention network to collaboratively learn user/item representations via a user-specific attention weight and a context-aware attention weight for a more accurate social recommendation. Fan et al.~\cite{GraphRec} proposed the GraphRec method, which can jointly capture interactions and opinions in the user-item interaction graph and consider heterogeneous strengths of social relations. Our work is orthogonal to the above works, i.e., the proposed method can be adopted to improve their performance if they are applied in the product recommendation problem of live stream E-Commerce.
\section{Problem Definition}

In this section, we define the problem of product recommendation in live streaming E-Commerce by modeling tripartite interaction information. Different from the conventional E-Commerce recommendation methods that only leverage the bipartite interactions between users and items, we have the tripartite relationship in the targeted  problem. The basic idea is to incorporate all the three kinds of bipartite interaction information, model the message propagation process and jointly learn the three tasks. To achieve this, it is desirable to first model the different kinds of interaction information as bipartite graphs and learn the unified representation to encode the tripartite entities. In this paper, we propose three bipartite graphs to model different interaction information, including user-item interaction graph, streamer-item interaction graph, and streamer-user interaction graph.

\textsc{Definition 1.} \textbf{\textit{(User-Item Graph)}} User-item graph, denoted as $\mathcal{G}_{(u,i)}=(\mathcal{U}\cup\mathcal{I},\mathcal{E}_{(u,i})$, captures the interaction information between users and items, and the interaction means users buy the corresponding items. $\mathcal{U}$ is a set of users and $\mathcal{I}$ is a set of items. $\mathcal{E}_{(u,i)}$ is the set of edges between users and items. We set the edge weight $w^{(u,i)_{jk}}$ as 1 if user $u_j$ buys item $i_k$, otherwise 0.
The user-item graph directly captures the interaction relationship between users and products, which is the essential information used by existing product recommendation approaches such as NGCF~\cite{NGCF}. 

Beyond this kind of interaction information, streamers bring another two kinds of interactions which can benefit the product recommendation. The streamer-item graph and streamer-user graph are defined as below:

\textsc{Definition 2.} \textbf{\textit{(Streamer-Item Graph)}} Streamer-item graph, denoted as $\mathcal{G}_{(s,i)}=(\mathcal{S}\cup\mathcal{I},\mathcal{E}_{(s,i})$, is a bipartite graph where $\mathcal{S}$ is a set of streamers and $\mathcal{I}$ is a set of items. $\mathcal{E}_{(s,i})$ is the set of edges between streamers and items. The edge weight $w^{(s,i)}_{jk}$ is set as 1 if the streamer $s_j$ sells item $i_k$, otherwise 0.

The user-item graph and streamer-item graph capture the interaction information between humans and products from different perspectives. The user-item graph encodes the \textbf{\textit{buying}} relation, while the stream-item graph encodes the \textbf{\textit{selling}} relation. To encode the \textbf{\textit{following}} relation between streamers and users, we introduce the streamer-user graph, which captures the interaction between humans.

\textsc{Definition 3.} \textbf{\textit{(Streamer-User Graph)}} Streamer-user graph, denoted as $\mathcal{G}_{(s,u)}=(\mathcal{S}\cup\mathcal{U},\mathcal{E}_{(s,u})$, is a bipartite graph where $\mathcal{S}$ is a set of streamers and $\mathcal{U}$ is a set of users. $\mathcal{E}_{(s,u})$ is the set of edges between streamers and users. The edge weight $w^{(s,u)}_{jk}$ is set as 1 if the streamer $s_j$ is followed by user $u_k$, otherwise 0.

These three types of graphs can be further integrated into one heterogeneous interaction graph.

\textsc{Definition 4.} \textbf{\textit{(Heterogeneous Tripartite Interaction Graph)}} The heterogeneous tripartite interaction graph is the combination of user-item, streamer-item, and streamer-user graphs constructed from three kinds of entities (user, item, and streamer) with their multi-view historical interaction data. It captures tripartite perspectives of interaction relationship between humans and products specifically in live stream E-Commerce.

Note that the definition of a heterogeneous tripartite interaction graph can be generalized to integrate other types of interaction graphs such as \textit{adding to cart} relation between users and items, and \textit{watching} relation between streamers and users. In this work we are using the three types of interaction graphs (buying, selling and following) as an illustrative example, which are much related to the product recommendation task.

Finally, we formally define the problem of product recommendation by modeling tripartite interaction information as follows:

\textsc{Definition 5.} \textbf{\textit{Product Recommendation with Tripartite Interaction Information}} Given a large collection of interaction data between users, products, and streamers in live stream E-Commerce scenarios, the problem of product recommendation with tripartite interaction information aims to learn and aggregate low-dimensional representations of tripartite nodes by embedding the heterogeneous interaction graph and then make recommendations based on the learned representations by multi-task prediction.
\section{Data Analysis}
\begin{figure}
  \centering
  \includegraphics[width=\linewidth]{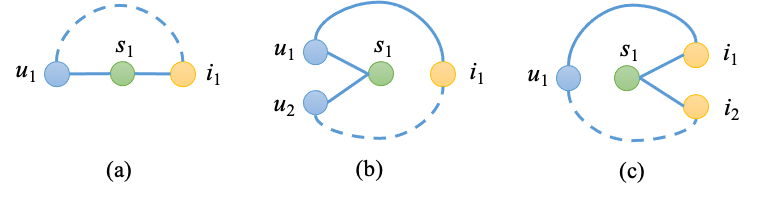}
  \vspace{-3ex}
  \caption{Examples of three different patterns that streamers could make an impact on users' purchase behavior. The solid lines mean the observed interaction behaviors and the dash lines mean the potential purchase behaviors.}
  \label{fig:sample}
  \vspace{-4ex}
\end{figure}

Before proposing the solution for product recommendation in live stream E-Commerce, in this section, we show the correlation between streamers and users' purchase by data analysis.
As shown in Figure~\ref{fig:sample}, there are three patterns among users and items by introducing streamers as mediums: 
(a) The streamer can connect users with items,
(b) The streamer can connect similar users,
(c) The streamer can connect similar products.
We put forward three questions corresponding to the three patterns respectively and present the 
detailed analysis to answer these questions.

\subsection{Relation Pattern between Users and Items}
This pattern means that the streamer 
acts as a bridge to connect the user and the item.
In this relationship, we aim to answer:
\begin{itemize}
    \item \textit{Whether there is a higher probability of purchase for the users facing the items sold by their following streamers?}
\end{itemize}
We conduct a simulation experiment by the Monte Carlo method~\cite{montecarlo} to calculate 
the probability of purchase for the users in two settings. 
The first setting $S_1$ is that the users are presented with items sold by their following streamers, while the second setting $S_2$ is that the users are presented with items which are not 
sold by their following streamers. 
We utilize set $P$ to denote the result in the first setting 
and set $Q$ to denote the result in the second setting.
The simulation process is described as :
\begin{enumerate}
    \item Initialize $P=\emptyset$ and $Q=\emptyset$.
    \item Sample a user and an item randomly.
    \item Set $t=1$ if the user purchases the item, otherwise set $t=0$.
    \item Put $t$ into $P$ for the first setting or $Q$ for the second setting.
    \item Repeat (2)-(4) for $N_{MC}$ times.
\end{enumerate}
Here, $N_{MC}$ is a large number of Monte Carlo simulations to ensure both $|P|$ and $|Q|$ are comparable to the total number of the interactions between users and items.
The probability of user purchasing items in the first setting $S_1$ can be calculated as follows:
\begin{equation}
    Prob_1 = \frac{P_{sum}}{|P|},
    \label{equa_p_1}
\end{equation}
where $P_{sum}$ denotes the sum of occurred purchases in set $P$. Similarly, We can also
get the probability of user purchase items in the second setting $S_2$ by replacing $P$ with $Q$ in the Equation~\ref{equa_p_1}. 
We run the experiments for 5 times and report the average results in Table~\ref{table:simulation}.
We observe that the user purchasing probability of items sold by his/her following streamers is approximately 5 times of those not sold by the streamers. The analysis result indicates that the User-Streamer-Item pattern can help to enhance the purchase probability between users and items, as shown in Figure~\ref{fig:sample}(a).

\subsection{Relation Pattern between Users}
This pattern indicates that the streamer 
acts as a bridge to connect two users.
In this pattern, we aim to answer:
\begin{itemize}
    \item \textit{When two users follow the same streamer, will their purchased items be more similar?}
\end{itemize}
Here, we utilize Cosine similarity and Jaccard similarity coefficient~\cite{Jaccard1912}
to measure the similarity 
between two item sets purchased by two users. 
For calculating the Cosine similarity $Cos_{jk}$ between user $u_j$ and $u_k$, 
we represent $u_j$ with a $N$-dimensional vector $v_j^u$, in which the $n$-th dimension is 1 if $u_j$ purchases the $n$-th item and 0 otherwise. The Cosine similarity $Cos_{jk}$ can be calculated:
\begin{equation}
    Cos_{jk} = \frac{\Braket{v_j^u,v_k^u}}{|v_j^u|*|v_k^u|}.
\end{equation}
The Jaccard similarity coefficient $Jac_{jk}$ between user $u_j$ and $u_k$ can be calculated as follows:
\begin{equation}
    Jac_{jk} = \frac{|\mathcal{I}_j \cap \mathcal{I}_k|}{|\mathcal{I}_j \cup \mathcal{I}_k|},
    \label{Jaccard}
\end{equation}
where $\mathcal{I}_j$ denotes the item set purchased by user $u_j$ and $\mathcal{I}_k$ denotes 
the item set purchased by user $u_k$.  
We use $\mathcal{U}_y^{pair}$ to denote the set of user pairs where two users follow the same streamer, and use 
$\mathcal{U}_n^{pair}$ to denote the set of user pairs where two users do not follow the same streamer. 
In order to study the distribution of the $Cos_{jk}$ and $Jac_{jk}$ in $\mathcal{U}_y^{pair}$ and $\mathcal{U}_n^{pair}$, 
we randomly sample multiple pairs from the set $\mathcal{U}_y^{pair}$ and $\mathcal{U}_n^{pair}$ respectively and 
calculate $Cos_{jk}$ and $Jac_{jk}$ for them. 
The statistical information (quantiles and mean value)
of the $Cos_{jk}$ and $Jac_{jk}$ for two sets are reported in Table~\ref{user_user}. 
We can learn that 
$Cos_{jk}$ and $Jac_{jk}$ in $\mathcal{U}_y^{pair}$ are significantly greater than those in $\mathcal{U}_n^{pair}$ 
on the quantiles and mean value. The result demonstrates that when two users follow the same streamer, they are more likely to purchase similar items. 
This experimental analysis indicates that the User-Streamer-User relationship can help to learn user similarities, which is beneficial to recommend items to similar users as illustrated in Figure~\ref{fig:sample}(b).

\begin{table}\small
	\begin{center}
		\caption{\textbf{The results of the simulation experiments.}}
		\vspace{-2ex}
		\small
		\begin{tabular}{ | c| c | }
			\hline
			\textbf{Setting}  & \textbf{Probability of purchase (avg $\pm$ std)}   \\ \hline
			$S_1$	& \textbf{3.35e-4}  $\pm$ \textbf{3.74e-7}  \\ \hline
			$S_2$	& 7.04e-5 $\pm$ 1.77e-7   \\ \hline
		\end{tabular}
		\label{table:simulation}
	\end{center}
	\vspace{-2ex}
\end{table}


\begin{table}
	\begin{center}\small
		\caption{\textbf{The quantiles and mean value of $Cos_{jk}$ and $Jac_{jk}$ in $\mathcal{U}_y^{pair}$ and $\mathcal{U}_n^{pair}$.}}
		\vspace{-2ex}
		\small
		\begin{tabular}{ | l | c | c|c|c|c|}
			\hline
			\textbf{Quantile Level} & \textbf{50\%} & \textbf{75\%} & \textbf{90\%} & \textbf{99\%}  & \textbf{Average}  \\ \hline
			$Cos_{jk}$ in $\mathcal{U}_n^{pair}$	& 0.0 	& 0.0	& 0.035
	& 0.22 & 0.012 \\ \hline
			$Cos_{jk}$ in $\mathcal{U}_y^{pair}$	& 0.0
	& \textbf{0.088}	& \textbf{0.177}	& \textbf{0.447} & \textbf{0.055} \\ \hline \hline
		$Jac_{jk}$ in $\mathcal{U}_n^{pair}$	& 0.0	& 0.0	&  0.0156 &
 0.10	& 0.0055  \\ \hline
			$Jac_{jk}$ in $\mathcal{U}_y^{pair}$	& 0.0 	& \textbf{0.0385}	& \textbf{0.0769}	& \textbf{0.25} & \textbf{0.0259} \\ \hline
		\end{tabular}
		\label{user_user}
	\end{center}
	\vspace{-2ex}
\end{table}

\begin{table}
	\begin{center}\small
		\caption{\textbf{The quantiles and mean value of $Cos_{jk}$ and $Jac_{jk}$ in $\mathcal{I}_y^{pair}$ and $\mathcal{I}_n^{pair}$.}}
		\vspace{-2ex}
		\small
		\begin{tabular}{ | l | c | c|c|c|c|}
			\hline
			\textbf{Quantile Level} & \textbf{98\%} & \textbf{99\%} & \textbf{99.9\%}  & \textbf{99.99\%} & \textbf{Average}  \\ \hline
			$Cos_{jk}$ in $\mathcal{I}_n^{pair}$	& 0.0  	& 0.0	& 0.031	& 0.10 & 8.7e-5 \\ \hline
			$Cos_{jk}$ in $\mathcal{I}_y^{pair}$	& 0.0 	& \textbf{0.01}	& \textbf{0.072}	& \textbf{0.18}  & \textbf{3.9e-4}\\  \hline \hline
			$Jac_{jk}$ in $\mathcal{I}_n^{pair}$	& 0.0 	& 0.0	& 0.011	& 0.044 & 3.5e-5 \\ \hline
			$Jac_{jk}$ in $\mathcal{I}_y^{pair}$	& 0.0 	& \textbf{0.003}	& \textbf{0.029}	& \textbf{0.098} & \textbf{1.59e-4} \\ \hline
		\end{tabular}
		\label{item_item}
	\end{center}
	\vspace{-2ex}
\end{table}

\subsection{Relation Pattern between Items}
This pattern means that the streamer 
acts as a bridge to connect two items.
In this relationship, we aim to answer:
\begin{itemize}
    \item \textit{When two items are sold by the same streamer, will the purchasers between them be more similar?}
\end{itemize}
We also adopt the Cosine similarity and the Jaccard similarity coefficient to represent the similarity 
between two user sets. 
For the Cosine similarity, we represent $i_j$ with a $M$-dimensional vector $v_j^i$ in which the $m$-th dimension is 1 if $i_j$ is purchased by the $m$-th user and 0 otherwise.
We use $\mathcal{U}_k$ to denote the set of users who purchased the item $i_k$ and use $\mathcal{U}_j$ to denote the set of users who purchased the item $i_j$. 
The Jaccard similarity coefficient $Jac_{kj}$ between $\mathcal{U}_k$ and $\mathcal{U}_j$ can also be calculated by Equation~\ref{Jaccard}.
Similarly, we use $\mathcal{I}_y^{pair}$ to denote the set of item pairs, where each pair represents that two items are sold by the same streamer. We utilize 
$\mathcal{I}_n^{pair}$ to denote the set of item pairs where two items are not sold by the same streamer.

The statistical information (quantiles and mean value)
of the $Cos_{jk}$ and $Jac_{jk}$ for $\mathcal{I}_y^{pair}$ and $\mathcal{I}_n^{pair}$ are shown in Table~\ref{item_item}.
Note that the quantity of an item being purchased by users is much larger than the quantity of a user purchasing items.
Hence, the similarities in $\mathcal{I}_y^{pair}$ and $\mathcal{I}_n^{pair}$ are much smaller than those in $\mathcal{U}_y^{pair}$ and $\mathcal{U}_n^{pair}$. Therefore, we set a larger quantile level in this analysis.
The results in Table~\ref{item_item} show that the Jaccard similarity coefficient and Cosine similarity in $\mathcal{I}_y^{pair}$ are greater than those in $\mathcal{I}_n^{pair}$ 
on both the quantiles and mean value, which indicates that two items sold by the same streamer are more similar.
As shown in Figure~\ref{fig:sample}(c), the analysis in this experiment indicates that the Item-Streamer-Item pattern can help to identify similar items for recommendation.

\section{Methodology}

\begin{figure*}
  \centering
  \includegraphics[width=0.8\linewidth]{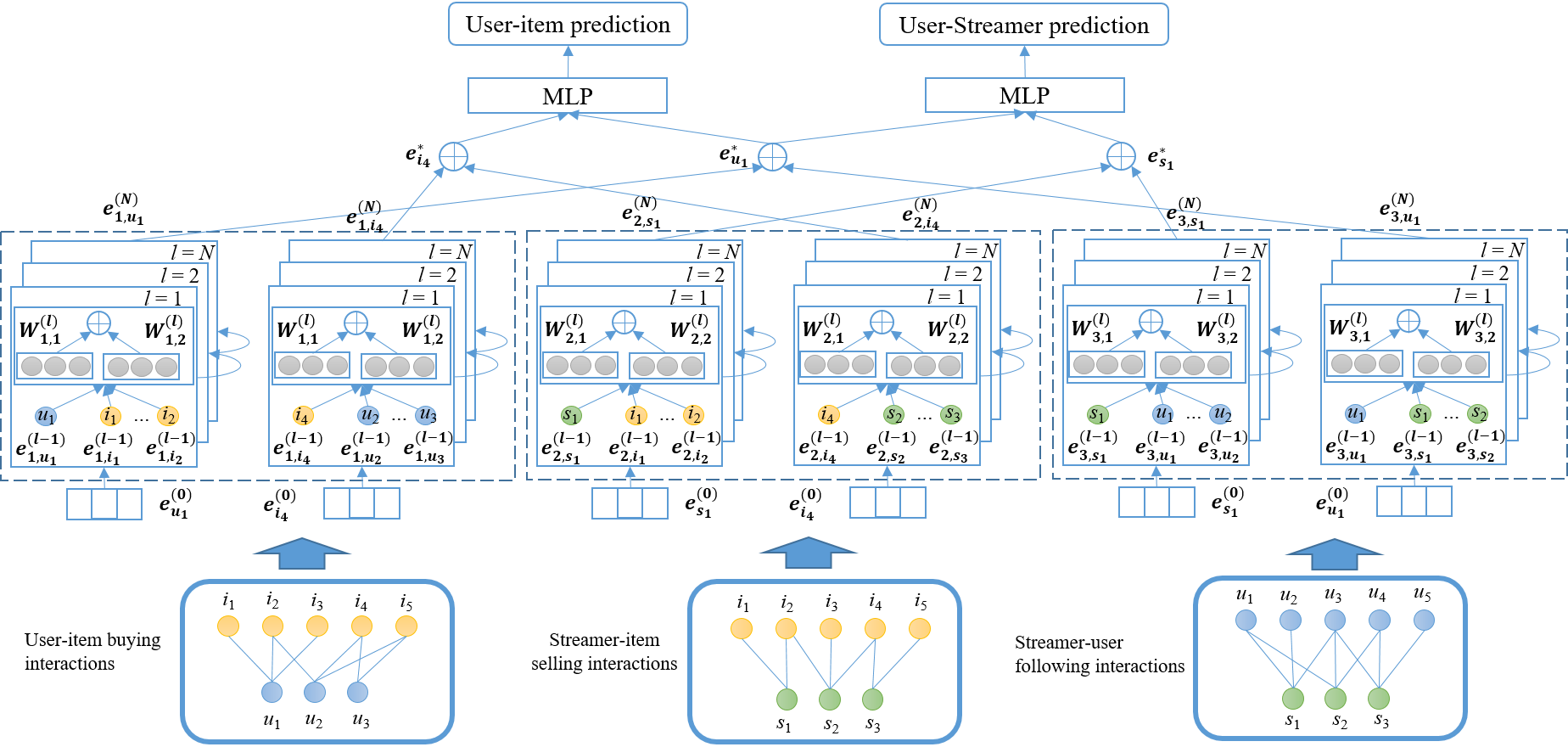}
  \caption{Overall architecture of LSEC-GNN framework (with GCN as the example aggregator).}
  \label{architecture}
\end{figure*}

Based on the previous analysis results, we introduce the details of the proposed LSEC-GNN framework, in which we perform graph representation learning by modeling tripartite interaction information from live stream E-Commerce to improve product recommendation. Our method first learns vector representations of tripartite nodes (users, items and streamers) by embedding the heterogeneous interaction graphs constructed from interaction data into a low dimensional space, and then the final node representations are concatenated from different bipartite graphs. At last, the final node representations are used to make predictions for multiple tasks. The overall architecture of LSEC-GNN is illustrated in figure \ref{architecture}.




\subsection{Bipartite Graph Embedding}

Since the heterogeneous tripartite interaction graph is composed of different bipartite graphs with the same node structure, here we do not distinguish them and introduce the consistent method. We first build the embedding lookup table for each node in the heterogeneous interaction graph as follows:
\begin{equation}
\begin{aligned}
	\mathbf{E}_U &= [ e^{(0)}_{u_1}, e^{(0)}_{u_2}, \cdots, e^{(0)}_{u_{|\mathcal{U}|}}]. \\
	\mathbf{E}_I &= [ e^{(0)}_{i_1}, e^{(0)}_{i_2}, \cdots, e^{(0)}_{i_{|\mathcal{I}|}}]. \\
	\mathbf{E}_S &= [ e^{(0)}_{s_1}, e^{(0)}_{s_2}, \cdots, e^{(0)}_{s_{|\mathcal{S}|}}].
\end{aligned}
\end{equation}
The lookup table is shared among bipartite graphs. Then in each bipartite graph, multiple embedding propagation layers are operated, which refines a node's representation by aggregating the embeddings of the interacted nodes. We do not limit the choice of propagation layers here and one can select any message-passing aggregator, such as GCN~\cite{GCN}, GAT~\cite{GAT}, NGCF~\cite{NGCF}, etc. Take GCN as the example, the $l$-th layer is updated as follows:
\begin{equation}
  H^{(l+1)}= \sigma\!\left(\tilde{D}^{-\frac{1}{2}} \tilde{A}\tilde{D}^{-\frac{1}{2}}H^{(l)} W^{(l)} \right) \,
\end{equation}
where $\tilde{A} = A + I_N$ is the adjacency matrix of a bipartite graph $\mathcal{G}$ with added self-connections.  $I_N$ denotes the identity matrix, $\tilde{D}_{ii} = \sum_j \tilde{A}_{ij}$, and $W^{(l)}$ is a layer-specific trainable weight matrix. $\sigma(\cdot)$ denotes the activation function. $H^{(l)}$ is the input embeddings in $l$-th layer.

Take the user-item bipartite graph of Figure~\ref{architecture} as an example, $e^{l-1}_{1,u_1}$ means the $u_1$'s embedding of the $(l-1)$-th layer in the $1$-st bipartite graph, $W^{(l)}_{1,2}$ means the $2$-nd $W$ in the $1$-st bipartite graph. Note that ideally our framework could model unlimited number of bipartite graphs. The input embedding $H^{(0)}$ and adjacency matrix $A$ can be formulated as follows
\begin{equation}
\begin{aligned}
    H^{(0)} &= [\mathbf{E}_U, \mathbf{E}_I]. \\
    A &= \begin{bmatrix}
0 & A_{ui} \\
A_{ui}^\intercal & 0
\end{bmatrix}.
\end{aligned}
\end{equation}
Here $A_{ui}$ is the user-item interaction matrix. At last after the graph operations, we can obtain the final representations of each node from corresponding bitpartite graphs.

\subsection{Interaction Graph Embedding}

In the heterogeneous tripartite interaction graph, where there are three bipartite graphs, each node of any type will have two embeddings from its belonging two bipartite graphs. After GNN operations, we can concatenate the outputs from each bipartite graph to generate the final representations of the tripartite nodes as follows:

\begin{equation}
\begin{aligned}
e^*_{i}=[e^{(N)}_{1,i},e^{(N)}_{2,i}] \\
e^*_{u}=[e^{(N)}_{1,u},e^{(N)}_{3,u}] \\
e^*_{s}=[e^{(N)}_{2,s},e^{(N)}_{3,s}]
\end{aligned}
\end{equation}

Then the final node representations are used for the following multiple predict tasks.


\subsection{Model Prediction}

After generating the unified embedding for each node from the heterogeneous tripartite interaction graph, we use the embeddings to make predictions about the existence of edges. Take a user $u$ as an example, we denote its embedding learned from \textit{user-item-buying} bipartite graph as $v_{u,\text{buy}}$ and the embedding from \textit{user-streamer-following} as $v_{u,\text{follow}}$. As the model training, the two embedding spaces should be close but the distance does exist. For item $i$, we denote its embedding learned from \textit{user-item-buying} bipartite graph as $v_{i,\text{buy}}$ and the embedding from \textit{streamer-item-selling} as $v_{i,\text{sell}}$. Now making a prediction based on the unified embedding of $u$ and $i$ will involve the interaction between totally three vector space, i.e., the vector spaces of three bipartite graphs. If we take the inner product of the unified embedding of $u$ and $i$, we will suffer from the issue that the inner product will be performed on two vectors from different vector spaces. Clearly, this cannot achieve optimal performance, because a simple inner product cannot capture the complex nonlinear relationships between different vector spaces.

To this end, we propose to concatenate the unified embedding of two nodes, and then use an MLP to learn the complex implicit interactions. For instance, with a two-layer MLP predictor, the score between a user $u$ and an item $i$ for recommendation task is computed as follows: 
\begin{equation}
\begin{aligned}
	\hat{y}_{ui} &= \sigma_2(\textbf{W}_2^T( \sigma_1(\textbf{W}_1^T ( [v_u \| v_i] ) + \textbf{b}_1)) + \textbf{b}_2),
\end{aligned}
\end{equation} where $\textbf{W}_x$, $\textbf{b}_x$ and $\sigma_x$ denote the weight matrix, bias vector and activation function for the $x$-th layer of MLP, respectively. $v_u$ and $v_i$ is the unified vector for user $u$ and item $i$, respectively.

As we have chosen to use MLP as the final predictor, we propose to apply another concatenating operation when combining embeddings from different bipartite graphs into a unified embedding for a node. In this way, both the interaction between a node's different embeddings and the interaction between different node's embeddings, can be learned in the final MLP predictor.

Empirically, such two-level concatenation and MLP predictor is the best combination strategy among all the methods we have tested. Therefore, in our experiments, we adopt this strategy.

\subsection{Multi-Task Optimization}

For model optimization, motivated by \cite{NCF}, we apply the simple but effective \textit{binary cross-entropy loss}, also known as \textit{log loss}. As our goal is product recommendation in  E-Commerce scenarios, we first optimize upon the \textit{user-item-buying} task and the objective function is formulated as follows:
\begin{equation}
\label{eqn:l_buy}
L_{\ \text{buy}} = -\sum_{(u,i)\in\mathcal{Y}\cup\mathcal{Y}^-} y_{ui}\log \hat{y}_{ui} + (1 - y_{ui}) \log (1 - \hat{y}_{ui}).
\end{equation}
Here $y_{ui}$ denotes the ground truth, in which the observed interactions are labelled as 1 and the unobserved interactions are labelled as 0. $\hat{y}_{ui}$ is the predicted score of $y_{ui}$. For each positive item that a user has interacted with, we randomly sample a fixed number of unobserved items as the negative instances.

However, with this single loss, the generated embeddings for streamers are not fully used and optimized. 
In live stream E-Commerce situations, the following behavior between a user and a live streamer often implies a kind of streamer's influence. To better optimize the embeddings for the streamers, we exploit an auxiliary task, which is to predict whether a user will follow a streamer. The loss function of this task is defined as follows:
\begin{equation}
\label{eqn:l_follow}
L_{\ \text{follow}} = -\sum_{(u,s)\in\mathcal{Y}\cup\mathcal{Y}^-} y_{us}\log \hat{y}_{us} + (1 - y_{us}) \log (1 - \hat{y}_{us}).
\end{equation}
The notation is similar to that of Equation~\ref{eqn:l_buy}. Meanwhile, the negative sampling strategy is also the same as the main task.

Finally, we take a linear combination of the above two loss functions as the final loss function as follows:
\begin{equation}
L = \alpha \cdot  L_{\ \text{buy}} + (1 - \alpha) \cdot L_{\ \text{follow}},
\end{equation}
where $\alpha$ is a hyper-parameter to be tuned.
\section{Experiments}

Since our target is to improve the product recommendation, two tasks (user-item prediction and user-streamer prediction) are involved and three kinds of interaction information are leveraged.



\subsection{Dataset Description}

As best as we know, there is no available open dataset for recommendation featuring E-Commerce streamers. we collect such data from the log of an E-Commerce website. The dataset ranges from December 1, 2020 to January 19, 2021. We randomly sample users, items, and streamers from the dataset, and further extract all the relevant mutual interactions as edges. We remove those who have less than 10 records from the dataset. Based on different sampling settings, we construct two datasets, i.e., \textbf{LSEC-Small} and \textbf{LSEC-Large}, for our experiments. The statistics of the two datasets is summarized in table \ref{table:data-stat}.
The interaction information includes the user-item buying data, user-streamer following data, and streamer-item selling data, and they are all binarized by erasing weights. We regard our experiments as the standard top-N recommendation problems. The proportion of training/validation/testing is set as 80/10/10\% by adopting chronological partition.


\begin{table}
    \caption{Statistics of the datasets.}
    \small
    \begin{tabular}{|l|c|c|}
        \hline
         \textbf{Term} & \textbf{LSEC-Small} & \textbf{LSEC-Large} \\
        \hline
         \# Item & 31,630 & 109,502 \\
        \hline
        \# User & 29,422 & 202,850 \\
        \hline
        \# Streamer & 4,633 & 7,395 \\
        \hline \hline
        \# Buying & 451,441 & 3,062,463 \\
        (Density) & (0.0485\%) & (0.0138\%) \\
        \hline
        \# Following  & 1,659,943 & 5,439,288 \\
        (Density) & (1.2178\%) & (0.3626\%) \\
        \hline
        \# Selling & 1,168,165 & 1,953,881 \\
        (Density) & (0.7972\%) & (0.2413\%) \\
        \hline
    \end{tabular}
    \label{table:data-stat}
    \vspace{-4ex}
\end{table}

\subsection{Experimental Settings}
\subsubsection{Evaluation Metrics}

To evaluate the product recommendation task, we refer to commonly-used metrics used in top-N recommendations~\cite{NGCF}, e.g. AUC, MRR, NDCG@\textit{K} and Recall@\textit{K}, where $K$ is set to $10$ and $50$.





\subsubsection{Baseline Models}

Since our proposed LSEC-GNN framework is general for various tripartite interaction graphs, different kinds of GNN blocks can be integrated as the message-passing aggregator, such as GCN and GAT. 
We choose one non-graph model and three graph-based models as our baselines:

\begin{itemize}
    \item \textbf{NCF}~\cite{NCF} includes three instantiations of neural CF: GMF, MLP, and the fusion of the former two, called NeuMF. In this work, NeuMF achieves the best performance, hence we choose it as the baseline.
    \item \textbf{GCN-RS} GCN~\cite{GCN} is a well-known graph-based model. In this work, a node's representation is updated by combining its original representation and the average of the neighbors' representations. It is originally proposed for node classification task and we adapt it for recommendation task. This method is denoted as GCN-RS.
    \item \textbf{LightGCN}~\cite{LightGCN} is a variant of GCN by removing feature transformation and nonlinear activation. For fair comparison, we replace the inner product predictor with MLP predictor to keep the same setting as our proposed methods. 
    \item \textbf{GAT-RS} GAT~\cite{GAT} is a variant of GCN by applying an attention mechanism to learn the weights for neighbors aggregation, instead of simply taking the average among them. We adapt it for recommendation task and name it GAT-RS.
\end{itemize}

We select graph-based models (i.e., GCN-RS, LightGCN, and GAT-RS), and use them as the aggregators for our LSEC-GNN framework. In this way, we propose three variants of our framework: \textbf{LSEC-GCN}, \textbf{LSEC-LightGCN}, and \textbf{LSEC-GAT}.


\subsubsection{Web-Scale Model Training}

Graph neural networks are typically conducted in a full-batch manner. 
Once the data get larger, GNN training will suffer from scalability issues. Motivated by GraphSAGE~\cite{GraphSAGE}, we apply the mini-batch training and neighborhood sampling on the graph. More specifically, given a batch of target nodes and edges, we build a subgraph by expanding neighbors according to the number of the GNN layers. Note that we conduct operations on a heterogeneous tripartite graph. Hence, it would become more complicated as we need create three subgraphs for each mini-batch and make sure there are as many common nodes as possible between two adjacent subgraphs. To reduce the receptive field size, we limit the number of neighbors expanded at each layer by a fixed threshold. Considering the different degree distributions of directional edges, we set the sampling threshold as the quantile $p$ of the degrees of a directional edge type. We implemented the aforementioned mini-batch training and neighbor sampling with the help of DGL~\cite{dgl}.

\subsubsection{Hyper-parameters Settings}

We optimize the models with Adam optimizer and set the learning rate to $0.0005$. The batch size is set to a relatively larger value $4096$ for speed as our experiments empirically show that it makes no difference. The GNN layer number $K$ for all graph-based models is set to $2$ after testing in the range of 1 to 4. The number of dimension for the embedding lookup table is set to $200$ and the input/output dimensions for the two GNN layers are $200 \rightarrow 128 \rightarrow 64$. For the neighbor sampling quantile $p$, we search in the range of \{$0.2$, $0.25$, $0.3$, $\cdots$, $1.0$ \} and finally set it as $0.9$. This can not only speed up the training process but also obtain better performance compared with taking all neighbors. The negative sampling ratio is set to $4$. We use a two-layer MLP predictor for each task. The loss coefficient $\alpha$ for multi-task training is set to $0.5$. We adopt hyperparameter tuning and early-stop mechanism on the validation set to choose the proper hyperparameters.


\subsection{Performance Comparison}

We repeat each experiment for 5 times and report the average metrics on the test set in table \ref{table:overall-result-small} and \ref{table:overall-result-large}. All experiments in this paper are conducted in this way.

\begin{table*}\small
    \caption{Comparison of the models on LSEC-Small dataset.}
    \vspace{-2ex}
    \begin{tabular}{lcccccc}
        \toprule
        Model & AUC & MRR & NDCG@10 & NDCG@50 & Recall@10 & Recall@50 \\
        \midrule
        NCF & 0.8103 & 0.1588 & 0.2392 & 0.3188 & 0.2832 & 0.5405 \\
        \midrule
        GCN-RS & 0.8440 & 0.1835 & 0.2862 & 0.3705 & 0.3363 & 0.6078 \\
LightGCN & 0.8483 & 0.1858 & 0.2895 & 0.3719 & 0.3374 & 0.6069 \\
GAT-RS & 0.8506 & 0.1828 & 0.2889 & 0.3742 & 0.3352 & 0.6183 \\
        \midrule
        LSEC-GCN & 0.8581 & \textbf{0.1924} & \textbf{0.3072} & \textbf{0.3869} & 0.3537  & 0.6205 \\
LSEC-LightGCN & \textbf{0.8641} & 0.1842 & 0.3022 & 0.3854 & \textbf{0.3615} & \textbf{0.6380} \\
LSEC-GAT & 0.8611 & 0.1873 & 0.3012 & 0.3867 & 0.3525 & 0.6375 \\
        \bottomrule
    \end{tabular}
    \label{table:overall-result-small}
    \vspace{-2ex}
\end{table*}

\begin{table*}\small
    \caption{Comparison of the models on LSEC-Large dataset.}
    \vspace{-2ex}
    \begin{tabular}{lcccccc}
        \toprule
        Model & AUC & MRR & NDCG@10 & NDCG@50 & Recall@10 & Recall@50 \\
        \midrule
     NCF & 0.8272 & 0.1767 & 0.2805 & 0.3608 & 0.3101 & 0.5633 \\
 \midrule
 GCN-RS & 0.8545 & 0.1988 & 0.3219 & 0.4091 & 0.3626 & 0.6365 \\
LightGCN & 0.8532 & 0.2001 & 0.3346 & \textbf{0.4184} & 0.3741 & 0.6363 \\
\midrule
 LSEC-GCN & 0.8482 & \textbf{0.2048} & \textbf{0.3374} & 0.4153 & 0.3735 & 0.6233 \\
LSEC-LightGCN & \textbf{0.8679} & 0.1981 & 0.3333 & 0.4157 & \textbf{0.3752} & \textbf{0.6427} \\
        \bottomrule
    \end{tabular}
    \label{table:overall-result-large}
    \vspace{-2ex}
\end{table*}

\begin{table*}\small
    \caption{Comparison of the LSEC-GCN model with different relations and tasks on LSEC-Small dataset.}
    \vspace{-2ex}
    \begin{tabular}{llcccccc}
        \toprule
        Relations & Tasks & AUC             & MRR             & NDCG@10         & NDCG@50         & Recall@10       & Recall@50       \\
        \midrule
        0         & 0     & 0.8440          & 0.1835          & 0.2862          & 0.3705          & 0.3363          & 0.6078          \\
        0, 1      & 0     & 0.8449          & 0.1864          & 0.2923          & 0.3731          & 0.3389          & 0.6036          \\
        0, 2      & 0     & 0.8453          & 0.1748          & 0.2749          & 0.3553          & 0.3270          & 0.5877          \\
        0, 1, 2   & 0     & 0.8547          & 0.1915          & 0.3057          & 0.3844          & 0.3509          & 0.6146          \\
        0, 1, 2   & 0, 1  & \textbf{0.8581} & \textbf{0.1924} & \textbf{0.3072} & \textbf{0.3869} & \textbf{0.3537} & \textbf{0.6205} \\
        \bottomrule
    \end{tabular}
    \vspace{-1ex}
    \label{table:ablation}
\end{table*}

We can obtain some observations from the experiment results. All graph-based models outperform the non-graph model NCF by a large margin. This is mainly because the graph model can capture the high-order connectivity between nodes. On both datasets, we find that LSEC-GNN based models are able to achieve the best performance on most of the metrics. This observation indicates that the tripartite interaction information is well captured and exploited, which is capable of improving product recommendation.


\subsection{Ablation Study on LSEC-GNN}

As our framework mainly consists of two parts: heterogeneous relations modeling and multi-task training, we will investigate their impacts on the performance. We use GCN as the example aggregator. We denote the three relations, \textit{user-item-buying}, \textit{user-streamer-following} and \textit{streamer-item-selling}, as $0$, $1$ and $2$, respectively. For the tasks, we denote the task of \textit{buying} (Eq.~\ref{eqn:l_buy}) as $0$ and the task of \textit{follow} (Eq.~\ref{eqn:l_follow}) as $1$. We first add relation type one by one to demonstrate the impact of each type of relations. Then we additionally consider the auxiliary task $1$ to show the influence of multi-task training. For instance, a single-relation and single-task GCN model for \textit{user-item-buying} recommendation can be expressed as \textit{relation 0, task 0}, and the final LSEC-GCN is marked as \textit{relations 0,1,2, tasks 0,1}. We run the experiments upon LSEC-Small dataset and the results are reported in table \ref{table:ablation}. We get the following observations:

\textbf{\textit{Impact of heterogeneous relations.}} As shown in the first four rows, adding relation $1$ (\textit{user-streamer-following}) can gain better overall scores compared with relation $0$ only. 
When both relation $1$ and $2$ are added, the performance can be further improved. This is because that when all relations are available, the interaction between different nodes can be well preserved and provide complementary information for each other, which leads to better node representations. 

\textbf{\textit{Impact of multi-task training}} The last two rows of the table show that the auxiliary task $1$ can make further improvement on the performance. This is because in the multi-task training setting, the embeddings of all three types of nodes are directly used in the loss calculation. When only the \textit{buying} task is optimized, the embeddings for streamers can still be updated as discussed in the Methodology section. But this kind of optimization is implicit. By directly optimizing the main task and the auxiliary task simultaneously, the complex tripartite interaction relationships can be learned in a more effective way. 

\subsection{Visualization of Streamer's Influence}

\begin{figure}
  \centering
  \includegraphics[width=0.8\linewidth]{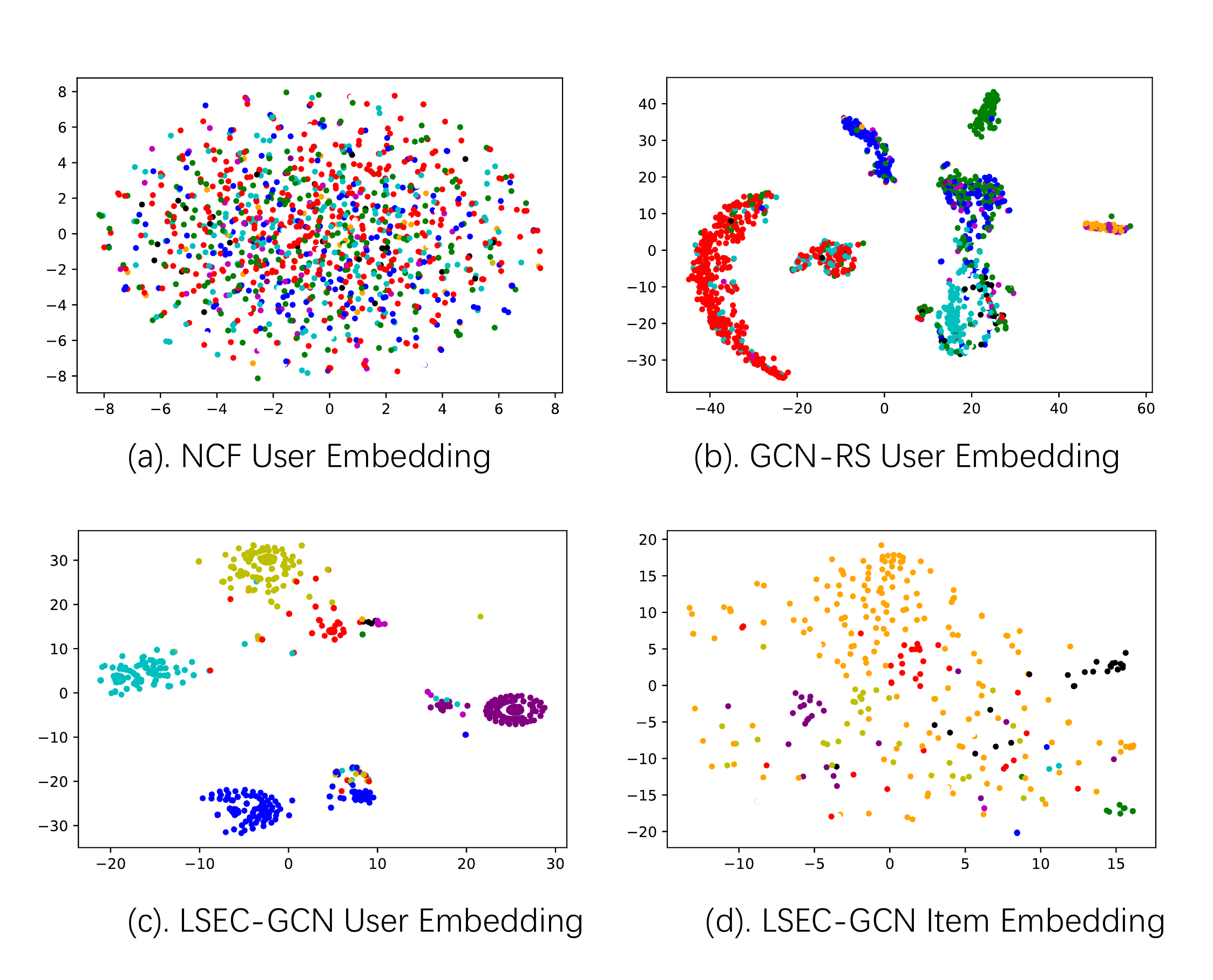}
  \vspace{-2ex}
  \caption{Visualization of user and item embeddings from different models with t-SNE.}
  \label{fig:tSNE}
  \vspace{-3ex}
\end{figure}

For further demonstrating our proposed method on the relation patterns, we utilize the t-SNE algorithm~\cite{tSNE} to visualize the \textbf{\textit{streamer's influence}}. To be more specific, we attempt to visualize the user embeddings and item embeddings learned by NCF, GCN-RS, LSEC-GCN and use the interacted streamers (\textit{following} interaction for a user and \textit{selling} interaction for an item) as labels. The users or items interacted with the same streamer are drawn in the same color. The visualization results are illustrated in figure~\ref{fig:tSNE}.

From Figure~\ref{fig:tSNE}(a), we observe that no clear patterns can be found in NCF user embedding. From Figure~\ref{fig:tSNE}(b) and Figure~\ref{fig:tSNE}(c), we can find that users are grouped into clusters, and the similar users in the same cluster are following the same streamer with a high probability. This suggests the superiority of GCN-RS and LSEC-GCN as graph-based methods over NCF, and implies the streamer's influence in grouping users. Comparing Figure~\ref{fig:tSNE}(b) and Figure~\ref{fig:tSNE}(c), our LSEC-GCN model exhibits a better clustering result in capturing the user's preference on streamers. As shown by Figure~\ref{fig:tSNE}(d), the item embedding from our LSEC-GCN model also shows a similar clustering result upon streamer’s influence on items, although not as obvious as that from those of user embeddings. In short, the visualization experiment not only shows the streamer’s influence in connecting among users and items, but also indicates our LSEC-GCN model's ability to model the streamer's influence.

\section{Conclusion}
In this paper, we investigate a novel problem of how to leverage the tripartite interaction information in live stream E-Commerce to improve product recommendation. We conduct data analysis on the collected real-world data. For the first time, we examine three interaction patterns among streamers, users, and products, and quantitatively study the streamer influence on users' purchase behavior. Based on the analysis results, we decompose the tripartite interaction information into several bipartite graphs, and propose a novel graph neural networks (LSEC-GNN) framework to aggregate the node representations of bipartite graphs. We conduct extensive experiments on two real-world live stream datasets, and the results demonstrate that our method is effective to improve the product recommendation performance. Moreover, we also present the visualizations of streamer influence.

\section{Acknowledgments}

Zhuoxuan Jiang is partially supported by Tencent Intelligent Media Platform Project. Qi Liu and Sanshi Lei Yu acknowledge the support of the USTC-JD joint lab. We thank the anonymous reviewers for their insightful comments.

\bibliographystyle{ACM-Reference-Format}
\bibliography{ref}

\end{document}